\documentclass[aps,prl,twocolumn,superscriptaddress,10pt]{revtex4-2}
\usepackage[utf8]{inputenc}
\usepackage{amsmath, bm}
\usepackage{cancel}
\usepackage{soul}
\usepackage{amsfonts}
\usepackage[normalem]{ulem}
\usepackage{mathtools}
\usepackage{physics}
\usepackage{dsfont}
\usepackage{verbatim}
\usepackage{color}
\usepackage[dvipsnames]{xcolor}
\usepackage[ruled,longend]{algorithm2e}
\usepackage{textgreek}
\usepackage{amssymb}
\usepackage{eucal}
\usepackage{tcolorbox}
\usepackage{wasysym}
\usepackage{float}
\usepackage{subfigure}
\usepackage{amsthm, amssymb,bbm}
 
\usepackage{url}
\usepackage{breakurl}

\begin{document}
\title{Liouvillian skin effect in quantum neural networks} 

\author{Antonio Sannia}
\affiliation{%
 Institute for Cross-Disciplinary Physics and Complex Systems (IFISC) UIB-CSIC, Campus Universitat Illes Balears, 07122, Palma de Mallorca, Spain.
}%

\author{Gian Luca Giorgi}
\affiliation{%
 Institute for Cross-Disciplinary Physics and Complex Systems (IFISC) UIB-CSIC, Campus Universitat Illes Balears, 07122, Palma de Mallorca, Spain.
}%

\author{Stefano Longhi}%
\affiliation{%
Dipartimento di Fisica, Politecnico di Milano, Piazza L. da Vinci 32, Milano I-20133, Italy}%
\affiliation{%
 Institute for Cross-Disciplinary Physics and Complex Systems (IFISC) UIB-CSIC, Campus Universitat Illes Balears, 07122, Palma de Mallorca, Spain.
}%

\author{Roberta Zambrini}%
\affiliation{%
 Institute for Cross-Disciplinary Physics and Complex Systems (IFISC) UIB-CSIC, Campus Universitat Illes Balears, 07122, Palma de Mallorca, Spain.
}%

\begin{abstract}
  In the field of dissipative systems, the non-Hermitian skin effect has generated significant interest due to its unexpected implications. A system is said to exhibit a skin effect if its properties are largely affected by the boundary conditions. Despite the burgeoning interest, the potential impact of this phenomenon on emerging quantum technologies remains unexplored. In this work, we address this gap by demonstrating that quantum neural networks can exhibit this behavior and that skin effects, beyond their fundamental interest, can also be exploited in computational tasks. Specifically, we show that the performance of a given complex network used as a quantum reservoir computer is dictated solely by the boundary conditions of a dissipative line within its architecture. The closure of one (edge) link is found to drastically change the performance in time-series processing, proving the possibility of exploiting skin effects for machine learning.
 \end{abstract}
\maketitle

\section{Introduction} 
A central assumption in statistical and condensed matter physics is that geometric boundaries and microscopic changes scarcely influence the global behavior of a physical system in the thermodynamic limit \cite{vanEnter1983,Hasler2006,prodan2005nearsightedness}. For example, in Ginzburg-Landau's theory of continuous phase transitions, symmetries play a key role, whereas other features, such as boundary conditions, are deemed negligible. Likewise, topological order and phase transitions in quantum matter \cite{RevModPhys.89.041004} have been regarded as boundary-independent phenomena. This paradigm has recently been severely challenged after the introduction of non-Hermitian (NH) topological systems and the discovery of the NH skin effect \cite{PhysRevLett.121.086803, PhysRevLett.121.026808,PhysRevB.97.121401,PhysRevX.8.031079,PhysRevLett.123.066404,PhysRevB.99.201103,PhysRevResearch.1.023013,torres2019perspective,PhysRevX.9.041015,PhysRevLett.124.086801,PhysRevLett.125.126402,Li2020,Helbig2020,xiao2020non,RevModPhys.93.015005, s42254-022-00516-5, Banerjee_2023, s11467-023-1309-z,UniversalSkin2022}. The NH skin effect, first introduced in  Ref.\cite{PhysRevLett.121.086803}, refers to the accumulation of eigenstates at the boundaries of a system, arising generally from asymmetric hopping amplitudes. This effect is distinct from conventional boundary phenomena in Hermitian systems, as it occurs even in the absence of external fields or disorder. The signature characteristics of the NH skin effect include the breakdown of the conventional bulk-boundary correspondence \cite{PhysRevX.8.031079,PhysRevLett.121.026808,PhysRevLett.121.086803,RevModPhys.93.015005}, non-Bloch phase transitions \cite{PhysRevLett.127.270602}, and transient self-acceleration \cite{s41467-024-48815-y}, to mention a few. These features are tied to the topology of the complex energy spectrum, which is non-Hermitian in nature. In dissipative quantum systems, a related phenomenon known as the Liouvillian skin effect can occur \cite{PhysRevLett.123.170401,Ueda_skin}. This effect involves the skin localization of eigenmodes of the Liouvillian superoperator, resulting in relaxation slowdown even in the absence of gap closing \cite{Ueda_skin}. The Liouvillian skin effect demonstrates the broader relevance of boundary phenomena in both non-Hermitian and dissipative contexts. Notably, it can also arise in the classical limit of pure incoherent hopping dynamics \cite{LonghiLSA2024}, further highlighting the importance of these effects across diverse quantum and classical physical systems. However, the emergence of the NH skin effect in systems beyond lattice configurations and its potential impact in the emerging areas of quantum technologies remain largely unexplored. Natural questions arise: moving beyond lattice configurations, can the boundary conditions still determine the behavior of complex quantum networks? Considering the interest of complex networks in the context of neuromorphic computing, can their performance in machine learning tasks be affected by NH skin effects?  

In this paper, we address both these questions going beyond classical settings, by introducing a class of quantum neural networks (QNNs), that present an irregular architecture, where a single specific network link determines their computational performances, even in the large size limit.

QNNs are machine learning models that translate classical artificial neural networks in the quantum domain, with the aim of achieving significant speed-ups in tackling computational problems compared to their classical counterparts \cite{QNN1, QNN2, QNN3, QNN4, QNN5}. Several QNN implementations and applications have been reported ranging from variational quantum algorithms with discrete \cite{QNN5, QNN_qcirc, QNN_qcirc2, Mitarai2018} and continuous variables \cite{QNN_CV}, to Quantum Boltzmann Machines \cite{QBM}, Quantum associative memories \cite{Ass_mem, Ass_mem2, Ass_mem3}, and Quantum Reservoir Computers (QRCs) \cite{opport, QRC_FN, QRC_rodrigo, QRC_Tran, QRC_Ang, govia2021quantum}.

In the following, we show the impact of the skin effect in QNNs, focusing on QRCs. Reservoir computing is a machine-learning algorithm designed for time series processing -a rather demanding task- with the advantage of easy training when compared with the challenges posed by deep or recurrent neural networks. The main idea behind this strategy is using the non-linear evolution of a dynamical system to process information, reducing training costs to a simple optimization at the output \cite{nakajima2021reservoir, van2017advances}. Physical (classical or quantum) platforms can be reservoir computers if a set of requirements are fulfilled. In particular, the state of a proper reservoir must depend only on the recent input history, as ensured by the so-called fading memory \cite{fading_memory} and echo state property  \cite{ESP}. Moreover, if a reservoir computer can be tuned for discriminating different time series \cite{separability} we say that it is universal because it can be used for reproducing any fading memory map, as reported both in classical and quantum settings \cite{sm, univ3,univ, univ2, Sannia2024}. Going beyond classical reservoirs, it is possible not only to process quantum inputs by avoiding complex embeddings and quantum tomography but also by exploiting the advantage offered by enlarged Hilbert space \cite{opport}. Complex bosonic networks have been considered as quantum reservoirs in both continuous and discrete variables \cite{Nokkala2022,univ2,Jorge.PhysRevApplied.20.014051,Llodr2022,Dudas2023}. Here we propose a novel design where directional dissipation plays a key role. We show that they can solve machine-learning tasks, and are also universal, if and only if the boundary conditions of the dissipative part of their architectures are not periodic.

\begin{figure}[t]
    \centering
     \includegraphics[width=\linewidth, keepaspectratio]{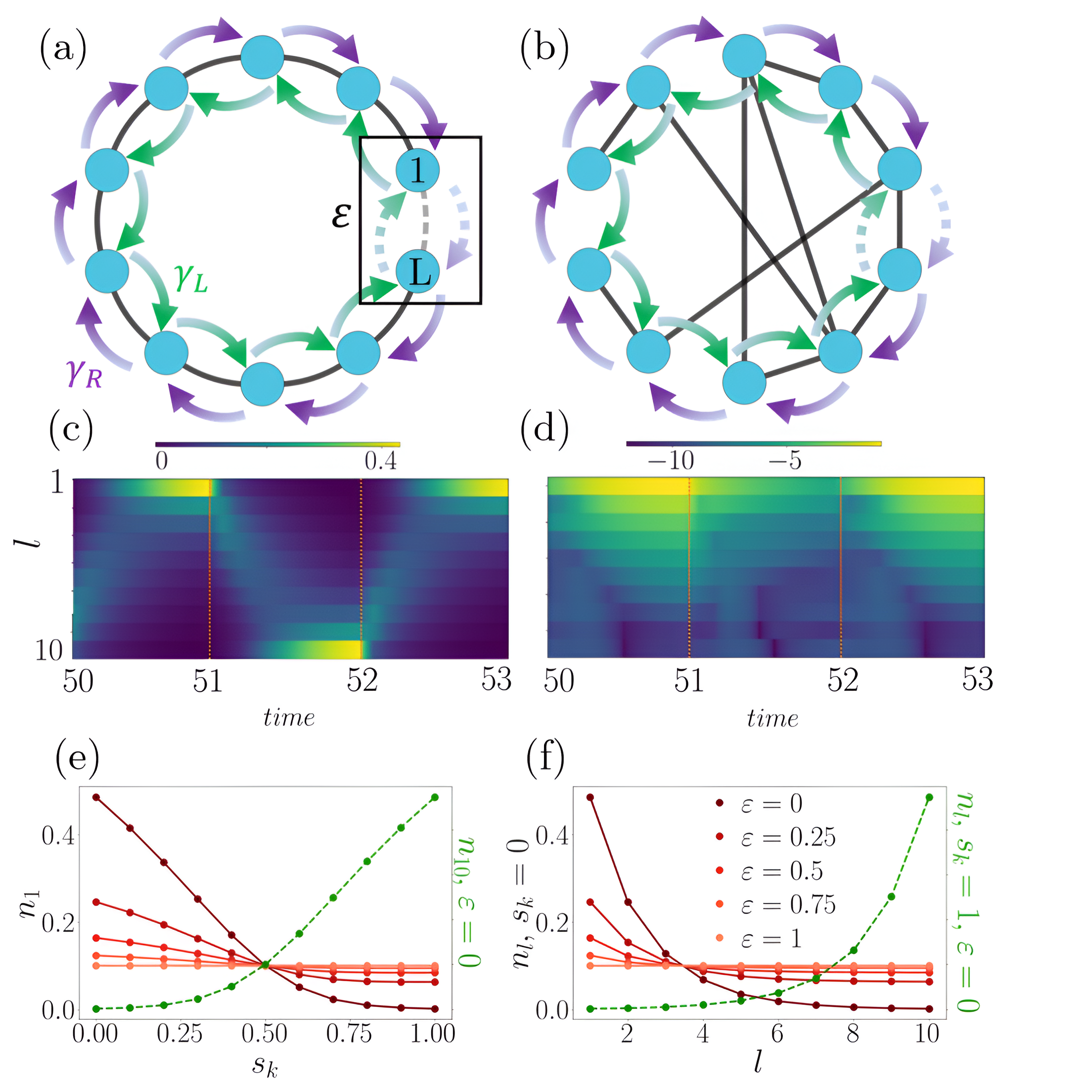}
    \caption{(a)-(b) Schematic of the two quantum networks considered in the analysis. Reservoir unitary (black line) and dissipative (green and purple lines) couplings (Eq. \ref{Eq:upd_rule}) for (a) chain and (b) complex network. Input is injected tuning the dissipative coupling strengths (see main text). (c)-(d) Time evolution of the populations ($\langle \hat{n}_l\rangle$) and state coherences between the first and the $l$ site  ($\log|\langle a_1 a_l^{\dagger} \rangle|$) for 3 input injections. (e)-(f) Populations for the stationary state of Eq. (\ref{Eq:upd_rule}) in the single excitation sector and for the network of panel (a) (linear chain). (e)  $\langle \hat{n}_1\rangle$ and $\langle \hat{n}_L\rangle$ as a function of the input  $s_k$ for different values of $\varepsilon$. (f) Populations for the extreme inputs  $s_k = 0, 1$ and increasing values of $\varepsilon$. Note that for $ \epsilon \neq 1$ the population tends to reside at one of the two edges of the lattice, a fingerprint of the Liouvillian skin effect. Hyperparameters values are: $L = 10$, $J =  W = \gamma =1$. }
    \label{fig:Fig1}
\end{figure}

\begin{figure*}[ht]
    \centering
   \includegraphics[width= \linewidth, keepaspectratio]{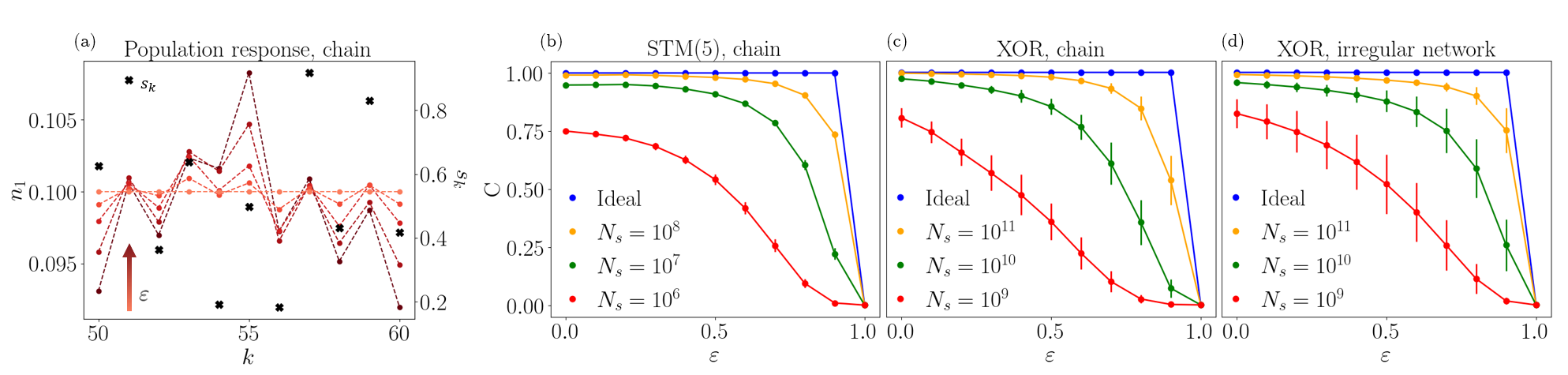}
    \caption{(a) First site population input response varying $\varepsilon$ for a single realization (chain network case). $50$ washout inputs, randomly extracted from the set $[0,1]$, have been injected to forget the initial condition. (b-d) Reservoir performances as a function of $\epsilon$ for the ideal case without shot noise, and in the presence of noise for a few increasing values of samples $N_s$. The error bars refer to one standard deviation computed over the considered realizations. (b) STM task with a delay $\tau = 5$ assuming $J = 1$, $W = 0.01$, $\Delta t = 1$, $\gamma = 0.1$, in the chain case. (c)-(d) XOR task for hyperparameters $J = 1$, $W = 1$, $\Delta t = 10$, $\gamma = 0.1$, considering both the chain and irregular network cases.}
    \label{fig:perf_shotnoise}
\end{figure*}

\section{Quantum reservoir models} 
We consider Markovian open quantum systems in which dissipation induces tunable asymmetric incoherent hoppings among various network nodes. The systems dynamics
is described by the Gorini-Kossakowski-Sudarshan-Lindblad (GKLS) master equation \cite{breuer2002theory,lindblad1976generators, gorini1976completely}:
\begin{equation} \label{eq:Linb}
\dot{\rho}=\mathcal{L}[\rho] \equiv -i[H,\rho]+\sum_{i}\gamma_{i} ( L_{i}\rho L_{i}^{\dagger}-\frac{1}{2} \{L_{i}^{\dagger}L_{i},\rho \})
\end{equation}
where $H$ is the system Hamiltonian, $\{\gamma_{i}\}$ are the damping rates that quantify the coupling strength with the external environment while $\{L_{i}\}$ are the jump operators, describing system-environment interaction. The Liouvillian $\mathcal{L}$, which generates the evolution of Eq. \eqref{eq:Linb}, comprises the unitary evolution part $\mathcal{U}[\rho]=-i[H,\rho]$ and the dissipative (Lindbladian) one $\mathcal{D}$ such that $\mathcal{L}=\mathcal{U}+\mathcal{D}$.

In our specific case, the Hamiltonian $H$ describes a complex (random) bosonic network  
\begin{eqnarray}\label{Eq:Hamiltonian}
    H &=& \sum_{l=1}^{L} w_{l} a^{\dagger}_l a_l + 
   \sum_{\substack{i,j=1 \\ i > j}}^{L} J_{i,j}( a_{i}a_{j}^{\dagger} +  a_{i}^{\dagger}a_{j}),
\end{eqnarray}
where $L$ is the network number of sites, $w_l$ are the disordered site energies, uniformly sampled from the fixed interval $[-W/2, W/2]$, $a_{l}$ and $a^{\dagger}_l$ are the creation and annihilation bosonic operators acting on the $l$-th site, and $J_{i,j}$ are the coherent hopping amplitudes (see Fig. \ref{fig:Fig1} (a-b)).

To create a proper quantum reservoir computer, it is essential to incorporate dissipation and input dependence into the reservoir evolution map \cite{Chen_2019}. In existing QRC models based on the Lindblad master equation, the input was injected into the reservoir through Hamiltonian driving \cite{Sannia2024,Dudas2023,govia2021quantum,QRC_Ang}. To unveil the major impact of skin effects in the quantum reservoir, in this paper, we follow a different strategy encoding the inputs into the Lindbladian term of evolution through asymmetric incoherent hopping amplitudes connecting the network nodes along a given closed path. Asymmetric hopping induces chirality, that is, preferential particle transport along one of the two circulation directions (left/right) of the path. By appropriate ordering of the site node numbers \{$l$\}, incoherent right/left particle hopping is realized by employing in Eq.(\ref{eq:Linb}) the jump operators $L_{R,l} = \sqrt{\gamma_{R}}a^{\dagger}_{l+1}a_l$ and $L_{L,l} = \sqrt{\gamma_{L}}a_{l+1}a^{\dagger}_l$ \cite{Ueda_skin}. Each input $s_k$, which is rescaled to vary in the range $[0,1]$, is encoded in the right/left hopping amplitudes by letting $\gamma_R = \gamma s_k$ and  $\gamma_L = \gamma (1 - s_k)$, where $\gamma$ measures the incoherent hopping strength. Note that for $s_k=1/2$ the left/right hopping amplitudes $\gamma_{L,R}$ are equal and particle transport is symmetric, whereas for $s_k=0,1$ one of them vanishes, resulting in unidirectional dissipative particle transport. 
Being the Lindbladian term non-local, boundary conditions should be properly specified. Let us define a Lindbladian with open boundary conditions $\mathcal{D}_{OB}(s_k)$ such that the jump operators entering in Eq.(\ref{eq:Linb}) are the $L_{R,l}$ and $L_{L,l}$ operators, with site index $l$ that varies from $1$ to $L-1$. Likewise, for periodic boundary conditions let us define the Lindbladian $\mathcal{D}_{PB}(s_k)$ such that the index $l$ appearing in the jump operators varies from $1$ to $L$, with the cyclic condition $l+L \rightarrow l$. Tunable boundary conditions are realized considering the following Lindbladian term in Eq.(\ref{eq:Linb})
\begin{align}
    \mathcal{D}(s_k, \varepsilon) =  \mathcal{D}_{OB}(s_k) + \varepsilon ( \mathcal{D}_{PB}(s_k) - \mathcal{D}_{OB}(s_k)),
\end{align}
where $\varepsilon$ is a dimensionless parameter, varying in the range $[0,1]$, that controls the connection strength at the boundary. The extreme limits $\varepsilon = 0$ and $\varepsilon = 1$ correspond to open and periodic boundary conditions, respectively. The asymmetry in hopping amplitudes is responsible for the appearance of the skin effect, as we will discuss below. A special and relevant system described by Eq. (\ref{Eq:Hamiltonian}) is a one-dimensional lattice with nearest-neighbor connections, corresponding to $J_{i,j} = J \delta_{i,j+1}$. This model can be physically implemented with ultracold atoms in an optical lattice by laser-assisted hopping with
spontaneous emission \cite{Ueda_skin}, where the asymmetric hopping amplitudes encoding the input data $s_k$ are controlled by the Rabi frequencies of the two control laser beams (see the Supplemental Material of Ref.\cite{Ueda_skin} for a detailed description of the model). In such a physical platform, an additional Lindbladian term responsible for an on-site dephasing $\mathcal{D}_d$ should be also considered. Such a dephasing term is generated by the jump operators $L_{d,l} = \sqrt{\gamma} a^{\dagger}_{l}a_l$ acting on all the lattice sites \cite{Ueda_skin}. Finally, the controllable on-site potential disorder in the optical lattice can be created using various techniques, such as optical speckle patterns, atomic mixtures, or inhomogeneous magnetic fields \cite{Fallani2008}. Other physical models that could exhibit controllable incoherent asymmetric hopping and disorder in a lattice topology include photonic quantum walks in synthetic lattices \cite{PhysRevLett.106.180403} and models of asymmetric simple exclusion processes in bosonic lattices \cite{10.21468/SciPostPhys.16.1.029}. Such systems could, in principle, be extended to two-dimensional geometries \cite{PengXue2024NatureCommun}.\\

To highlight the role of skin effect, we consider as illustrative examples two network configurations, which are depicted in Figs. \ref{fig:Fig1} (a) and (b). The first one [Fig.\ref{fig:Fig1} (a)] is a simple linear chain (lattice topology) with connections only between nearest neighbors. Here, the boundary connection strength is identical in both the Lindbladian and the Hamiltonian part of the dynamics, ($J_{L,1} = \varepsilon J$). The second configuration involves a more generic network topology, with coherent hoppings identically equal to $J$ between random pairs of nodes [Fig. \ref{fig:Fig1} (b)].

To summarize, at each time step \textit{k} the updating rule of the reservoir state $\rho_k$ is described by the following master equation
\begin{align}\label{Eq:upd_rule}
   \rho_{k+1} = e^{\mathcal{L}(s_{k+1}, \varepsilon) \Delta t} \rho_{k} = e^{\left(\mathcal{U}+\mathcal{D}(s_{k+1},\varepsilon) + \mathcal{D}_d\right) \Delta t}\rho_k,
\end{align}
which defines the basic dynamical evolution of the dissipative quantum system and QRC operation. As usual, the reservoir processes the input information and, in this case, this is described by Eq. \ref{Eq:upd_rule}. Subsequently, a set of expectation values of observables is optimized to achieve the target task \cite{sm}.

\section{Skin effect and QRC performances analysis} 
A key feature of the reservoir model defined by Eq.(\ref{Eq:upd_rule}) is that boundary conditions on the Lindbladian term govern the spectrum and eigenstate localization properties of the Liouvillian superoperator, a hallmark of the Liouvillian skin effect \cite{PhysRevLett.123.170401,Ueda_skin}. This, in turn, has a profound impact on the QRC system's performance. To illustrate such behavior, let us first observe that the evolution generated by the Liouvillian $\mathcal{L}(s_{k}, \varepsilon)$ preserves the total number of bosons $N_{\rm tot}=\sum_{i = 1}^{L} a_l^{\dagger}a_l$, which constitutes a strong symmetry, and, as proved in Theorem 1 of the Supplemental material \cite{sm} if we restrict its action to a Hilbert space sector with a fixed number of bosons then its stationary state is unique. As demonstrated in \cite{Sannia2024}, the uniqueness of the stationary state for fixed input is a sufficient condition for a reservoir to exhibit the echo state property (ESP) \cite{sm}. In our specific case, however, if the reservoir state is not an eigenstate of $N_{\rm tot}$ --number of excitations not fixed-- different initial states will generally converge to distinct final states after being subjected to the same input sequence, thereby violating the ESP. This is due to the fact that the operator corresponding to the total number of excitations constitutes a strong symmetry of the Liouvillian superoperator. Therefore, if we consider two different initial states
belonging to different excitation sectors, each of them will converge towards the corresponding steady state of its own sector. Consequently, to ensure the validity of the ESP in our analysis, we must assume that the initial state of the reservoir always belongs to a specific excitation sector.

However, the presence of ESP alone is not sufficient to have a useful reservoir. As shown in Ref. \cite{Rodrigo_PRE}, if the stationary state of $\mathcal{L}(s_{k}, \varepsilon)$ is the same for any value of $s_k$ then the state of the reservoir will always converge to this fixed state. 
Consequently, all possible observables we can measure will be independent of the particular time series that is injected, rendering them useless for extracting any information from it. Interestingly, in this system, the presence of open or closed boundaries is responsible for the proper reservoir operation. As proved in Theorems 2 and 3 of the Supplemental material \cite{sm}, in the case of periodic boundary conditions in the Lindbladian, i.e. $\varepsilon = 1$, the Liouvillian stationary state is always the fully mixed one, independently on $s_k$, and, consequently, the reservoir state will converge to it losing all the time series information. Conversely, when $\varepsilon \neq 1$, the stationary state depends on the injected input $s_k$. As proved in Theorem 3 of the Supplemental material \cite{sm}, for any $\varepsilon \neq 1$, two different input values cannot correspond to the same stationary state. Such a property is illustrated in panels (e,f) of Fig.1, which show the numerically computed population distribution at equilibrium versus $s_k$ and for a few values of $\epsilon$. Only when $\epsilon=1 $, i.e. under periodic boundary conditions, the steady state is the fully mixed state, independent of $s_k$, and the reservoir computing task is expected to fail. This means that skin modes, originating by breaking periodic boundaries in the system, are essential to realize QRC. Moreover, this property also allows us to conclude that the reservoirs described by Eq.(\ref{Eq:upd_rule}) can discriminate a general pair of different time series, forming a universal class of reservoir computers \cite{sm}.

The performances of the reservoir model are illustrated by considering the popular RC benchmark temporal tasks for the two different topologies shown in Fig. \ref{fig:Fig1} (a)-(b): chain and irregular network. In Fig. \ref{fig:perf_shotnoise}(a) we show the dynamical response of the internal state of the reservoir to the input series injection. If $\varepsilon \neq 1$, the reservoir reacts to the input at each time step (browner points), also displaying some memory effects. Indeed similar input values, such as the 5th and the 7th in the figure, produce different responses due to the previous input history. Otherwise, for periodic boundary conditions (light orange points) the system converges to a steady state that is uncorrelated to the injected series, as predicted by our theory.

In the following numerical simulations, we will always consider the initial reservoir state to be a random one belonging to the single excitation sector Hilbert space, fixing $L = 10$ network sites. To evaluate the performances, we employed the capacity $C=\mathrm{cov}^{2}(\textbf{y},\bar{\textbf{y}})/\sigma(\textbf{y})^{2}\sigma(\bar{\textbf{y}})^{2}$, where $\mathrm{cov}(\cdot)$ is the covariance, $\sigma(\cdot)$ is the standard deviation, and $\textbf{y}$ and $\bar{\textbf{y}}$ are, respectively, the target and prediction time series. The possible values that $C$ can take range from $C=0$ (the reservoir is completely useless for the task at hand) and $C=1$ (target series perfectly reproduced). To fully evaluate the potential of the entire reservoir, the output layer is constructed taking a linear combination of all the density matrix elements. In order to take into account the statistical uncertainty associated with quantum measurements, we will assume that $N_s$ samples are available \cite{sm}.

A typical benchmark test is used to evaluate the linear memory: the short-term memory (STM) task  \cite{STM}. Considering the input $s_k$ to be randomly sampled from the interval $[0,1]$ at each time step, the task consists of reproducing at the output past input values up to a certain delay. More precisely, given a delay value $\tau$, the target series $y_k$ to be reproduced is given by the relation: $y_{k}=s_{k-\tau}$. 

In Fig. \ref{fig:perf_shotnoise} (b), we show the capacity values for the STM task, fixing a delay $\tau = 5$, letting $\varepsilon$ vary, and also changing the chain network topology. The results indicate that avoiding periodic boundary conditions is essential for achieving significant performance. In the ideal scenario without shot noise, achievable when $N_s \sim 10^{15}$, the capacity exhibits discontinuous behavior as a function of $\epsilon$,
while, decreasing $N_s$, the performance decline becomes smoother. This behavior is related to the magnitude of the density matrix element changes during the dynamics. In fact, as it can be seen in Fig. \ref{fig:perf_shotnoise} (a), as $\varepsilon$ increases, the fluctuations of the reservoir state over time tend to become smaller and, consequently, more samples are needed to properly resolve the dynamics. This result explains why, in a realistic scenario, having fewer samples and a stronger coupling connection $\varepsilon$ degrades the performance. For the STM task, the same qualitative results are observed even with the irregular network topology.

The performances of the quantum reservoirs have also been tested considering the XOR task \cite{PC}, which, unlike the STM task, requires the ability to process inputs non-linearly. In this case, the $s_k$ values are discrete and randomly taken from the set $\{0,1\}$ while the desired target is $y_{k}=\sum_{j=1}^{2}s_{k-j} \mod 2$. The obtained results are shown in panels (c,d) of Fig. \ref{fig:perf_shotnoise}. As for the STM task, the reservoir performances are greatly affected by boundary conditions for both chain and irregular network topologies. A similar dependence on $N_s$ is also observed.

\begin{figure}[t]
\centering
\includegraphics[width= \linewidth, keepaspectratio]{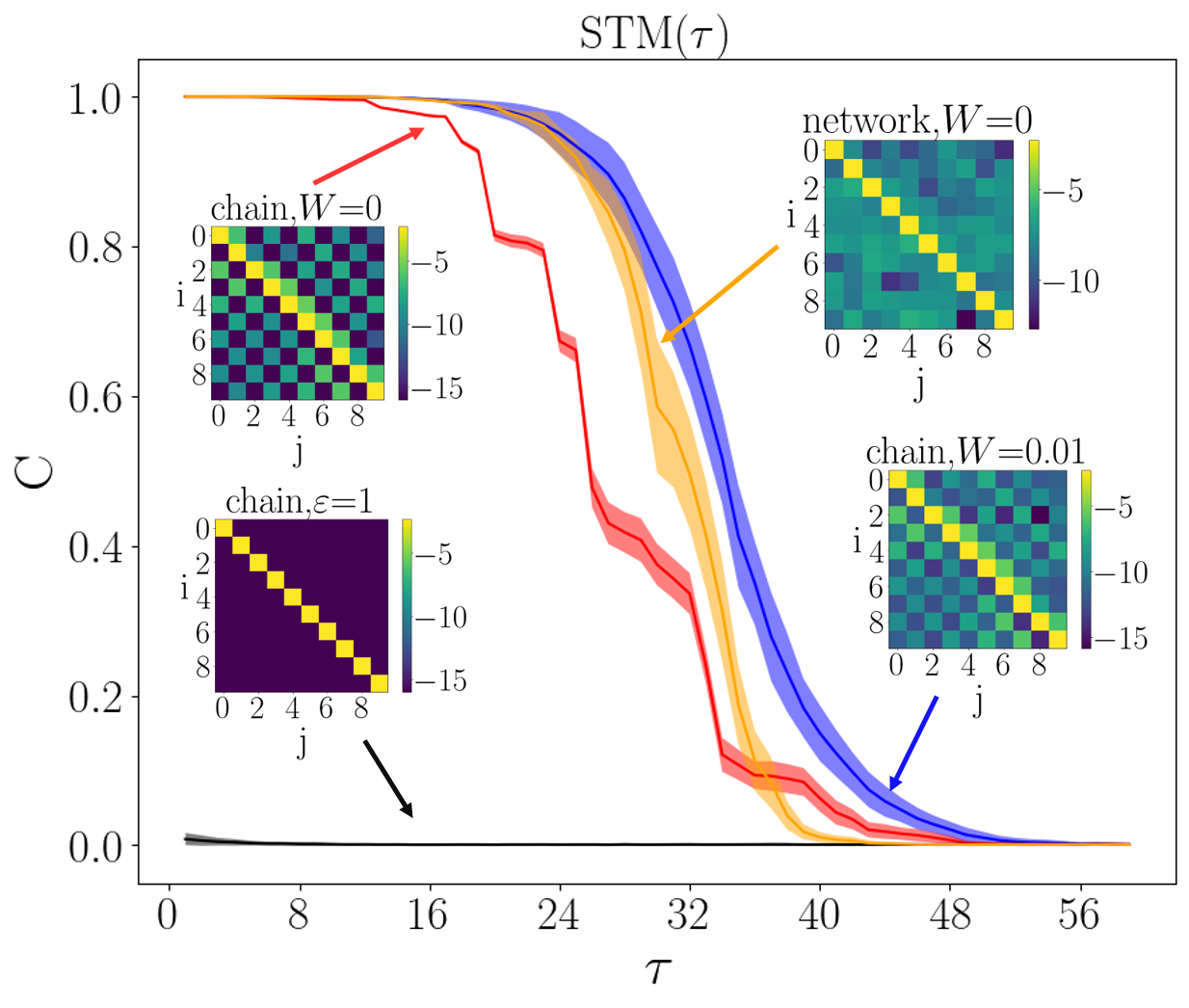}
    \caption{Short-term memory task performance, as a function of $\tau$, for the two network topologies shown in Fig.1(a,b). We considered the chain topology with $W = 0$ (fully ordered case), with $W = 0.01$ (on-site disorder), and the irregular network topology with $W = 0$ (network disorder). Hyperparameter values are $L=10$, $\varepsilon = 1$, $J = 1$, $\Delta t = 1$ and $\gamma = 0.1$ (ideal case). For comparison, we also show the results for the chain network with $\varepsilon = 1$, where the STM performance is fully degraded. The shadow regions cover one standard deviation over the various realizations considered \cite{sm}. For each case, we show the density matrix patterns after $300$ random input values injected on a random initial condition. The values in the patterns with $i \geq j$ refer to $\log_{10}|\Re{\rho_{ij}}|$ while for $i < j$ they refer to $\log_{10}|\Im{\rho_{ij}}|$.}
    \label{fig:disord}
\end{figure}

\section{Role of disorder} A source of disorder is a common ingredient in quantum reservoir computing as it helps to eliminate possible system symmetries,
responsible for reducing the degrees of freedom that can be extracted from the dynamics \cite{QRC_FN,QRC_rodrigo,QRC_Tran,QRC_Ang,univ2,Nokkala2022,Sannia2024,Jorge.PhysRevApplied.20.014051,Llodr2022,Chen_2019}. We study the effect of disorder either through the on-site potential, arising from randomly choosing the $w_l$ values in Eq. \ref{Eq:Hamiltonian}, or by making irregular the network topology. In Fig. \ref{fig:disord}, we show the effects of these two sources of disorder on the STM task performances; for reference, the performances of the fully ordered reservoirs are also reported. The numerical simulations clearly indicate better performances in the presence of disorder, which thus provides a resource in QRC. A simple explanation of this result can be gained by comparing the patterns of the density matrix elements after the same number of inputs, in the presence or absence of disorder; see Fig. \ref{fig:disord}. Clearly, in the disordered case the number of non-negligible matrix density elements is significantly higher, implying richer dynamics and a higher sensitivity of the stationary state on the input $s_k$, as expected. Interestingly, our numerics also suggest that these two different ways of introducing the disorder give comparable improvements.

\section{Discussion and conclusion}
In conclusion, we unveiled the universality of skin effects in open quantum systems beyond lattice configurations and demonstrated how boundary-driven phenomena can provide a fascinating resource in quantum neural networks and machine learning tasks. Specifically, we presented quantum reservoir computers based on bosonic networks, in which the input is injected through a controlled asymmetric particle transport in the evolution. We found a surprising sensitivity in computational tasks within quantum networks. Even in irregular network architectures, adding or removing a single link on a dissipative path can significantly impact the performance. Our results unravel the potential impact of skin effects in quantum networks, which could be of major relevance in emerging areas of quantum science and technologies. In our manuscript, we focused primarily on the Liouvillian skin effect on a closed loop created within QNNs, however, a broader class of skin effects could be possible in the context of QNNs, depending on the specific topology, interactions, and dynamical rules governing the system. For instance, in higher-dimensional QNN architectures, the NH skin effect could manifest as multidimensional localization patterns, where eigenstates accumulate along corners, edges, or faces \cite{UniversalSkin2022}. These patterns depend on the topology of the network and the interplay between dissipative and coherent dynamics. Additionally, QNNs with time-dependent non-Hermitian terms might exhibit transient skin effects, where localization properties evolve dynamically, potentially enhancing or suppressing information flow in certain network pathways. Other kinds of NH skin effect, such a the critical skin effect leading to scale-free localization \cite{Li2020}, could be considered as well. These effects might reveal unique pathways for dissipative learning or noise-enhanced phenomena, which could be intriguing topics for future investigations.

\section*{ACKNOWLEDGMENTS}
We acknowledge the Spanish State Research Agency, through the Mar\'ia de Maeztu project CEX2021-001164-M funded by the MCIU/AEI/10.13039/501100011033, through the COQUSY project PID2022-140506NB-C21 and -C22 funded by MCIU/AEI/10.13039/501100011033, MINECO through the QUANTUM SPAIN project, and EU through the RTRP - NextGenerationEU within the framework of the Digital Spain 2025 Agenda. The CSIC Interdisciplinary Thematic Platform (PTI+) on Quantum Technologies in Spain (QTEP+) is also acknowledged. GLG is funded by the Spanish  Ministerio de Educaci\'on y Formaci\'on Profesional/Ministerio de Universidades and co-funded by the University of the Balearic Islands through the Beatriz Galindo program (BG20/00085).
The project that gave rise to these results received the support of a fellowship from the ”la Caixa” Foundation (ID 100010434). The fellowship code is LCF/BQ/DI23/11990081.
\bibliography{bibliography}

\onecolumngrid
\section*{Supplemental Material}
\setcounter{section}{0}

\section{Quantum reservoir computing}\label{Sec:QRC}

In this section, we provide some formal results concerning general concepts and proprieties of a quantum reservoir computing device, such as
updating rules, echo state property, and fading memory.

\subsection{Quantum reservoir updating rule}
Given an input sequence that we want to analyze $\{\dots, s_{k-1}, s_{k}, s_{k+1}, \dots\}$, where $k$ is a generic time-step, let $\rho_k$ be the quantum reservoir state, $\Phi(s_k)$ the input-dependent quantum channel that describes the reservoir evolution, and $\{\mathcal{O}_i\}_i$ the set of observables that we measure at each time step.

The updating rule of the reservoir state is governed by the relation:
\begin{align*}
    \rho_{k+1} = \Phi(s_{k+1}) \rho_{k}.
\end{align*}
The reservoir output is a sequence of numbers $\{\dots, x_{k-1}, x_{k}, x_{k+1}, \dots\}$ such that 
\begin{align*}
    x_{k} = \alpha^{\textit{out}}_0 + \sum_i \alpha^{\textit{out}}_i \Tr{\rho_{k} \mathcal{O}_i}
\end{align*}
where the coefficients of the linear combination $\{\alpha^{\textit{out}}_i\}_i$ are optimized with a linear regression procedure, depending on the specific task of interest \cite{Cucchi2022}. 
\subsection{Echo state property}\label{Sec:Sup_echo}
The echo state property means that the quantum reservoir state dependence on the initial condition tends to decrease over input injections \cite{ESP}. 

Assuming that the reservoir states belong to a finite-dimensional Hilbert space, we formalize this idea by injecting a generic left infinite and time-ordered input sequence to two general initial conditions. Let $\{\cdots, s_{-1},s_{0} \}$ be the input series (where the 0-th time step corresponds to the last input) and let $\rho_1$ and $\rho_2$ be two initial reservoir states; we say that the model has the echo state property if the following pointwise limit holds
\begin{equation}
 \label{Eq:Echo}
    \lim_{k \to -\infty} \| \prod_{i=k}^{0} \Phi(s_k) \left( \rho_1 - \rho_2 \right) \| = 0, 
\end{equation}
where $\| \cdot \|$ is a generic matrix norm (we do not need to specify the norm since all the norms are equivalent in finite-dimensional vector spaces). The echo state property is satisfied when the quantum channel $\Phi(s_k)$ exhibits strict contractivity: $||\Phi(s_k)(\rho_1-\rho_2)||\leq r||\rho_1-\rho_2||$ for every $s_k$ and any pair of states $\rho_1$ and $\rho_2$, with $0\leq r<1$. 

The set of open quantum systems described by Eq. 4 of the main text evolves according to a Markovian map. Consequently, strict contractivity is ensured if the Liouvillian $\mathcal{L}(s_k, \varepsilon)$ which generates the dynamics admits only one stationary state and the evolution time $\Delta t$ at each time step is long enough. In particular, it is always possible to find a mixing time $\Delta t_{mix}$ such that if $\Delta t \geq \Delta t_{mix}$ then the echo state property holds \cite{Sannia2024}. 

\subsection{Fading memory}

When the reservoir state only depends on the recent input history we say that the fading memory property holds \cite{fading_memory}. Formally, this concept is translated to a continuous condition that a reservoir model has to satisfy. Indicating with $K^{-}([0,1])$ the space of the left infinite input sequences whose values are limited to the range $[0,1]$, we start by defining a norm on this particular space. Given a generic null sequence $\bm{w}=\{w_k\}_{k\geq 0}$, where $\lim_{k\rightarrow\infty}w_k=0$, the corresponding norm of a generic series $\bm{s} \in K^{-}([0,1])$ will be:
\begin{equation*}
    \| \bm{s}\|_{\bm{w}}=\sup_{k \in \mathbb{Z}^{-}}|s_{k}|w_{-k},
\end{equation*}
where $s_{k}$ and $w_{-k}$ are respectively elements of $\bm{s}$ and $\bm{w}$, and $\mathbb{Z}^{-}$ indicates the set of negative integers.

We say that a reservoir model presents the fading memory property if, after injecting an input sequence $\bm{s}$ to one generic initial condition, the expectation values of the observables that define its output layer $\biggl\{\Tr{\rho_{0} \mathcal{O}_i}\biggr\}_{i}$ are continuous functions of $\bm{s}$ under the introduced norm. As proved in \cite{Sannia2024}, for a quantum reservoir model described by a master equation, like the one in Eq. 4 of the main text, the presence of the echo state property directly implies fading memory if the Liouvillian that generates the dynamics continuously depends on the input. In our case study, it can be trivially verified that $\mathcal{L}(s_k, \varepsilon)$ is a continuous function $s_k$, implying that the fading memory property holds for the quantum reservoir model proposed, if $\Delta t \geq \Delta t_{mix}$. 

\section{Properties of the models} 

In this section, we prove that the performances of the quantum reservoir models, described by Eq. 4 of the main text, exhibit a skin effect. In particular, we first show the uniqueness of the $\mathcal{L}(s_k, \varepsilon)$ stationary state, and, afterward, we discuss its dependence on the boundary conditions of the dissipative loop. In particular, we prove that, when they are periodic, for any input value the stationary state is always the fully mixed state, implying that the reservoir models cannot be used for solving machine learning tasks. On the contrary, we also show that in the complementary case, $\varepsilon \neq 1$, two different inputs will always correspond to different stationary states. In this case, as we will discuss, the quantum reservoir models, in addition to being able to process temporal information, are also universal.

\subsection{Uniqueness of the stationary state}
We now prove that the stationary state of the Liouvillian $\mathcal{L}(s_k, \varepsilon)$ is unique in any sector of the Hilbert space with a fixed number of excitations. The fact that the reservoir state belongs to this space is a condition that can be easily obtained. In fact, if the initial state has a defined total number of bosons $N_{\rm tot}=\sum_{l=1}^{L}n_l$ ($L$ is the number of the network sites), then this value is conserved during the dynamics: $\mathcal{L}^{\dagger}(s_k, \varepsilon)\sum_{i = 1}^{L} a_l^{\dagger}a_l=0$.

Building on the recent insights from Ref. \cite{Zhang2024} about the Evans theorem \cite{Evans1977}, we first demonstrate that the only strong symmetry of $\mathcal{L}(s_k, \varepsilon)$ is the total number of excitations. Subsequently, we restrict our analysis to the corresponding symmetry sector to establish the uniqueness of its stationary state.

Let $I$ be a set of indexes and let us denote by $\{A_i, i \in I \}^{'}$ the set of the commutants of the $A_i$ operators, i.e. the set of all the linear operators that commute with all the $A_i$ ones. We can now enunciate the following result proven in \cite{Evans1977}:
\vspace{0.25cm} 

\textbf{Theorem (Evans):} \textit{Let $\mathcal{M}$ be a von Neumann algebra on the Hilbert space $\mathbb{H}$, which is globally invariant under the semi-group $e^{tL}$. The superoperator $L$ is assumed of the form}
\begin{equation*}
L(X)=V(X)+K^{\dagger}\,X+X\,K,\quad \mbox{with}\,X\in\mathcal{B}(\mathbb{H}),
\end{equation*} 
\textit{and}
\begin{equation*}
V(X)=\int_{\Omega}A(\omega)^{\dagger}XA(\omega)\mbox{d}\mu(\omega),\quad K=iH-\frac{1}{2}V(\mathbbm{1}),
\end{equation*}
\textit{where $\mathcal{B}(\mathbb{H})$ is the set of bounded operators acting on the Hilbert space, $H$ is a Hermitian operator referring to the system Hamiltonian and $(\Omega, \mu)$ is $\sigma$-finite measure space.}

\textit{If $T_t$ is the induced W*-dynamical semigroup on M, then the algebra $\mathcal{M}(T)=\{X\in M:T_{t}(X^{\dagger}X)=X^{\dagger}X,\,T_{t}(X)=X,\,\forall t\geq 0\}$ is equal to $\mathcal{M}\cap \{A(\omega),\,K\}'$. Thus $T_t$ is irreducible if and only if $\mathcal{M}\cap \{A(\omega),\,A(\omega)^{\dagger},\,H\}'=\mathbb{C}\mathbbm{1}$.}

\vspace{0.25cm} 
The definition of irreducibility presented in this statement corresponds to the absence of strong symmetries, as shown in \cite{Zhang2024}. Moreover, as explained in \cite{Nigro2019}, this general result can be adapted to our Liouvillian, in which we have a discrete set of jump operators, simply noticing that $V(X) = \sum_i L_i^{\dagger}XL_i$ in our specific case. 

We now observe that a sufficient condition for the absence of strong symmetries is that the set of the commutants of the jump operator is given by the operators proportional to the identity. 

We now enunciate the following Lemma:

\vspace{0.25cm} 
\textbf{Lemma 1:} \textit{Let the operators $L_i$ be the jump operators of the generator $\mathcal{L}(s_k, \varepsilon)$ that describe the reservoir evolution of Eq. 4 of the main text. Furthermore, let $N_{b}$ be the fixed number of bosons which defines the excitation sector of the Hilbert space of interest, $\mathbb{H}_{N_{b}}$. Then, the set of commutants of the jump operators, intersected with the set of the bounded operators acting on the considered Hilbert space $\mathcal{B}(\mathbb{H}_{N_b})$, is trivial: $\mathcal{B}(\mathbb{H}_{N_b}) \cap \{L_i\}'=\mathbb{C}\mathbbm{1}_{N_{b}}$, where $\mathbbm{1}_{N_{b}}$ is identity operator of \hspace{0.1cm}$\mathbb{H}_{N_{b}}$. Consequently, all the Liouvillians that share the same Lindbladian of $\mathcal{L}(s_k, \varepsilon)$ does not have any strong symmetry if they act on states with a fixed number of excitations and the Hamiltonian is compatible with the conservation of the total number of bosons.}
\vspace{0.25cm} 

Proof:

Let $\mathcal{O}$ be a generic element of $\mathcal{B}(\mathbb{H}_{N_b})$. In general, we can expand it in the Fock basis:
\begin{equation*}
   \mathcal{O} = \sum_{\substack{n_1, \dots, n_L; m_1, \dots, m_L\\ \sum_j n_j = \sum_j m_j = N_{exc}}} \alpha_{n_1, \dots, n_L; m_1, \dots, m_L} \ket{n_1, \dots, n_L} \bra{m_1, \dots, m_L}.
\end{equation*}

Imposing that $\mathcal{O}$ commutes with all the jump operators of $\mathcal{L}(s_k, \varepsilon)$ will allow us to calculate the elements of the set of commutants. Let us first notice that the commutation condition with the jump operators of $\mathcal{D}_d$ (the on-site dephasing Lindbladian), which are the lattice site number operators $\{n_l\}_{l=1}^{L}$ (unless a trivial multiplicative constant that we can disregard), significantly simplifies the form of $\mathcal{O}$. In fact, it is straightforward to check that if $[\mathcal{O}, n_l] = 0$ $\forall l$ then $\mathcal{O}$ is a diagonal operator: $\mathcal{O} = \sum_{\substack{n_1, \dots, n_L\\ \sum_j n_j = N_{b}}} c_{n_1, \dots, n_L} \ket{n_1, \dots, n_L} \bra{n_1, \dots, n_L}$.

We now impose the commutation condition between $\mathcal{O}$ and the operators $\{a_la_{l+1}^{\dagger}\}_{l=1}^{L-1}$. In particular, we want the following relationship to be valid: $[a_la_{l+1}^{\dagger},\mathcal{O}]= 0$  $\forall l$. For the sake of definiteness, let us fix the number of bosons on all the sites, except for two adjacent sites, say $l$ and $l+1$ (say $\sum_{j\neq l,l+1}n_j=\bar{N}_l$). We can schematically rewrite the operator $\mathcal{O}=\sum_x c_x \ket{x}\bra{x}$, where $\ket{x}$ are all possible combinations of $\ket{n_l,n_{l+1}}$ compatible with the constrain $n_l+n_{l+1}=N_b-{\bar N}_l$. In this basis,  $a_la_{l+1}^{\dagger}$ will induce transition between pairs of such states: $a_la_{l+1}^{\dagger}=\sum_{x,y}f_{x,y} \ket{x}\bra{y}$, where $f_{x,y}$ takes into account the statistics of bosonic operators. It is easy to show that $\mathcal{O}a_la_{l+1}^{\dagger}=\sum_{x,y}c_x f_{x,y} \ket{x}\bra{y}$, while $a_la_{l+1}^{\dagger}\mathcal{O}=\sum_{x,y}c_y f_{x,y} \ket{x}\bra{y}$. We can conclude that $[a_la_{l+1}^{\dagger},\mathcal{O}]= 0$ if and only if $c_x=c_y\; \forall x,y$. By repeating the same argument for all pairs of nearest neighbors, we conclude that all the coefficients in the diagonal expansion of $\mathcal{O}$ must be identical. $\blacksquare$

It is worth noting that the proof of Lemma 1 could have also been developed in the same way by imposing the commutation of $\mathcal{O}$ with the reflected set of operators $\{a_l^{\dagger}a_{l+1}\}_{l=1}^{L-1}$. This observation allows us to conclude that the strong symmetries are not present for any input value $s_k$, even in the extreme cases where $s_k = 0$ or $s_k = 1$. It is also important to pinpoint that we did not have to take into account the operators $a_1^{\dagger}a_L$ and $a_L^{\dagger}a_1$ that refer to the Lindbladian action on the lattice borders. As a consequence, this result holds even in the other extreme case of open boundary conditions when $\varepsilon = 0$.

We now link the absence of strong symmetries with the uniqueness of the stationary state enunciating

\vspace{0.25cm} 
\textbf{Theorem 1:} \textit{Let $\mathcal{L}(s_k, \varepsilon)$ be the Liouvillian of Eq. 4 of the main text and let $N_{b}$ be a fixed number of bosons which defines a Hilbert space excitation sector $\mathbb{H}_{N_{b}}$, then $\mathcal{L}(s_k, \varepsilon)$, acting only on a $\mathbb{H}_{N_{b}}$ subspace, admits only one stationary state.}
\vspace{0.25cm} 

Proof:

Indicating again with $L_i$ the $\mathcal{L}(s_k, \varepsilon)$ jump operators, as explained in \cite{Zhang2024}, Frigerio's second theorem \cite{Frigerio1978} proves that the Liouvillian stationary state is unique when $\{L_i\}'$ is trivial and $span\{L_i\}$ is an adjoint set. As a consequence of Lemma 1, $\mathcal{L}(s_k, \varepsilon)$ directly fulfills these two conditions when $s_k$ belongs to open set $(0,1)$. The extension of this result, in the extreme cases in which $s_k = 0$ or $s_k = 1$, is ensured by the fact that the Hamiltonian realizations that may violate the uniqueness of stationary state have zero probability of being sampled (see Appendix A 1  of  Ref. \cite{Minganti2018}). $\blacksquare$

As already mentioned, the uniqueness of the stationary state, for any value of the input $s_k$, ensures that we can always define a mixing time $\Delta t_{mix}$ such that if $\Delta t \geq \Delta t_{mix}$ then the echo state property holds. In \cite{Ueda_skin}, the $\Delta t_{mix}$ scaling of the Liouvillian $\mathcal{L}(s_k, \varepsilon)$, referred to Eq. 4 of the main text, has been numerically computed, restricting its action on the single excitation sector and considering the cases of open and periodic boundary conditions, empirically finding that $\Delta t_{mix} = O(L^2)$. We believe that, in the most general case in which $0<\varepsilon<1$ and the number of bosons is $N_b$, it is reasonable to assume that $\Delta t_{mix} = O(\text{poly}(N_b, L))$. 

\subsection{Stationary state for periodic boundary conditions}

We will now show another important property of the Liouvillian $\mathcal{L}(s_k, \varepsilon)$:

\vspace{0.25cm} 
\textbf{Theorem 2:} \textit{Let $\mathcal{L}(s_k, \varepsilon)$ be the Liouvillian generator of Eq. 4 of the main text, let $N_{b}$ be the fixed number of bosons which defines the excitation sector of the Hilbert space of interest $\mathbb{H}_{N_{b}}$ then the unique stationary state of $\mathcal{L}(s_k, \varepsilon)$ in the case of periodic boundary conditions, i.e. $\varepsilon = 1$, is the fully mixed state: $\mathbb{I}_{N_{b}}/\text{dim}(\mathbb{H}_{N_{b}})$.}
\vspace{0.25cm} 

Proof:

Firstly, we observe that the uniqueness of the stationary state is ensured by Theorem 1 and, consequently, showing that $\mathcal{L}(s_k,1)[\mathbb{I}_{N_{b}}] = 0$ is sufficient for proving the entire statement of Theorem 2. Writing in extended form the Liouvillian action, we have:
\begin{eqnarray*}
    \mathcal{L}(s_k,1)[\mathbb{I}_{N_{b}}] &=& -i[H, \mathbb{I}_{N_{b}}] + \gamma s_k \Bigl( \sum_{l=1}^{L-1} a_{l}a_{l+1}^{\dagger}\mathbb{I}_{N_{b}}a_{l}^{\dagger}a_{l+1} + a_{L}a_{1}^{\dagger}\mathbb{I}_{N_{b}}a_{L}^{\dagger}a_{1} - \frac{1}{2}\sum_{l=1}^{L-1} \{a^{\dagger}_l a_l (1+a^{\dagger}_{l+1} a_{l+1}), \mathbb{I}_{N_{b}} \} \\ 
    &-& \frac{1}{2} \{a^{\dagger}_L a_L (1+a^{\dagger}_{1} a_{1}), \mathbb{I}_{N_{b}} \} \Bigr) + \gamma (1 - s_k) \Bigl( \sum_{l=1}^{L-1} a_{l}^{\dagger}a_{l+1}\mathbb{I}_{N_{b}}a_{l}a_{l+1}^{\dagger} + a_{L}^{\dagger}a_{1}\mathbb{I}_{N_{b}}a_{L}a_{1}^{\dagger} \\
    &-& \frac{1}{2}\sum_{l=1}^{L-1} \{a^{\dagger}_{l+1} a_{l+1} (1+a^{\dagger}_{l} a_{l}), \mathbb{I}_{N_{b}} \} - \frac{1}{2} \{a^{\dagger}_1 a_1 (1+a^{\dagger}_{L} a_{L}), \mathbb{I}_{N_{b}} \} \Bigr) + \mathcal{D}_d[\mathbb{I}_{N_{b}}] \\
    &\vcentcolon=&-i[H, \mathbb{I}_{N_{b}}] + \gamma s_k \mathcal{D}_R[\mathbb{I}_{N_{b}}] + \gamma (1-s_k) \mathcal{D}_L[\mathbb{I}_{N_{b}}] + \mathcal{D}_d[\mathbb{I}_{N_{b}}].
\end{eqnarray*}
Noticing that, trivially, it holds that $-i[H, \mathbb{I}_{N_{b}}] = \mathcal{D}_d[\mathbb{I}_{N_{b}}] = 0$ we focus on the remaining terms. Expanding $\mathbb{I}_{N_{b}}$ in the Fock basis, 
\begin{eqnarray*}
   \mathbb{I}_{N_{b}} = \sum_{\substack{n_1, \dots, n_L\\ \sum_j n_j = N_{b}}} \ket{n_1, \dots, n_L} \bra{n_1, \dots, n_L},
\end{eqnarray*}
it readily follows that 
\begin{eqnarray*}
    \mathcal{D}_R[\mathbb{I}_{N_{b}}] =  \sum_{\substack{n_1, \dots, n_L\\ \sum_j n_j = N_{b}}} \Bigr(\sum_{l=1}^{L}(n_l + 1)n_{l+1} + (n_L + 1)n_1 - \sum_{l=1}^{L}n_l(n_{l+1}+1) -  n_L(n_1 + 1)\Bigl) \ket{n_1, \dots, n_L} \bra{n_1, \dots, n_L} = 0.
\end{eqnarray*}
In the same way, it is possible to show that $\mathcal{D}_L[\mathbb{I}_{N_{b}}] = 0$, concluding that $\mathbb{I}_{N_{b}}/\text{dim}(\mathbb{H}_{N_{b}})$ is the unique stationary state of $\mathcal{L}(s_k,1)$. $\blacksquare$

It is interesting to notice that the proof of Theorem 2 only requires that $H$ conserves the total number of bosons to make Theorem 1 apply and, therefore, the result can be trivially generalized to a wider class of Liouvillians. 
We also remark that, even in this case, the result is true for any input value $s_k$. The result of Theorem 2 has the physical interpretation that, in the case of periodic boundary conditions, the asymmetry in the incoherent left/right hopping amplitudes governed by the input amplitudes $s_k$ does not basically influence the long-time dynamical behavior of the system. 

\subsection{Skin effect in the performances}

We can now show that, for the considered quantum reservoir, the possibility of processing the information of a time series strongly depends on the boundary conditions, implying a skin effect. Reformulating Theorem 2 of \cite{Rodrigo_PRE} for this case of study, we prove the following  

\vspace{0.25cm} 
\textbf{Theorem 3:} \textit{Given a proper quantum reservoir, whose state belongs to the $N_b$ bosons excitation sector, which respects the echo state property and evolves according to Eq. 4 of the main text, then, in the case of periodic boundary conditions, for any input series injected into the system the reservoir state will always converge to a fully mixed state $\mathbb{I}_{N_{b}}/\text{dim}(\mathbb{H}_{N_{b}})$, being useless for extracting any kind of information from it.}
\vspace{0.25cm} 

Proof:

We start observing that, by definition of echo property in Eq. \ref{Eq:Echo}, the state to which the reservoir tends to converge is independent of its particular initial condition. Consequently, given an infinite and time-ordered input sequence $\{\cdots, s_{-1},s_{0} \}$, we can easily compute the general final reservoir state $\rho_{f}$ by assuming as an initial state the fully mixed state $\mathbb{I}_{N_{b}}/\text{dim}(\mathbb{H}_{N_{b}})$. From Theorem 2, one has
\begin{equation*}
    e^{\mathcal{L}(s_k, 1) \Delta t} \mathbb{I}_{N_{b}}/\text{dim}(\mathbb{H}_{N_{b}}) = \mathbb{I}_{N_{b}}/\text{dim}(\mathbb{H}_{N_{b}}) \quad \forall  s_k
\end{equation*}
and, consequently, we can determine $\rho_{f}$ using a direct calculation:
\begin{equation*}
    \rho_{f} = \lim_{k \to -\infty} \prod_{i=k}^{0} e^{\mathcal{L}(s_k, 1)\Delta t} \rho_{in} =  \lim_{k \to -\infty} \prod_{i=k}^{0} e^{\mathcal{L}(s_k, 1)\Delta t} \mathbb{I}/L = \mathbb{I}/L
\end{equation*}
where $\rho_{in}$ is a generic initial condition. $\blacksquare$

From Theorem 3 we conclude that working under open ($\varepsilon=0$) or semi-open ($ 0<\varepsilon <1$) boundary conditions is a necessary requirement for the quantum reservoir model to perform any machine learning task. 

Contrary to the case of $\varepsilon = 1$, when $0 \leq \varepsilon < 1$, the $\mathcal{L}(s_k,\varepsilon)$ stationary states will have an explicit dependence on $s_k$. Going into more detail, we prove this stronger statement:

\vspace{0.25cm} 
\textbf{Theorem 4:} \textit{Let $\mathcal{L}(s_k, \varepsilon)$ be a Liouvillian generator, which is described by Eq. 4 of the main text, acting on the $N_{b}$ bosons excitation sector, then for every fixed $0 \leq \varepsilon < 1$, for different input values $s_k$ the corresponding unique stationary states of the $\mathcal{L}(s_k, \varepsilon)$ will be in turn different.}
\vspace{0.25cm} 

Proof:

As already done, we apply Theorem 1 to conclude that the stationary state of $\mathcal{L}(s_k, \varepsilon)$ is always unique regardless of the particular value of $s_k$ and $\varepsilon$. We now write $\mathcal{L}(s_k, \varepsilon)$ in a more concise form which will be useful for the next calculations:
\begin{equation*}
    \mathcal{L}(s_k, \varepsilon)[\cdot] = -i[H, \cdot] + \gamma s_k \mathcal{D}_R(\varepsilon)[\cdot] + \gamma (1 - s_k) \mathcal{D}_L(\varepsilon)[\cdot] + \mathcal{D}_d [\cdot]
\end{equation*}
where $\mathcal{D}_R(\varepsilon)$ and $\mathcal{D}_R(\varepsilon)$ are the parts of Lindbladian that contain respectively the jumps operators of the form $a_{l}a_{l+1}^{\dagger}$ and $a_{l}^{\dagger}a_{l+1}$.

Let's denote with $s_1$ and $s_2$ two generic different input values, then if the corresponding Liouvillians $\mathcal{L}(s_1, \varepsilon)$ and $\mathcal{L}(s_2, \varepsilon)$ admit the same stationary state $\rho_{ss}$ it must hold that: $\bigl(\mathcal{L}(s_1, \varepsilon) - \mathcal{L}(s_2, \varepsilon)\bigr)\rho_{ss} = 0$. From the previous equation, we easily arrive at a more specific condition that a generic stationary state $\rho_{ss}$ shared between two different inputs must fulfill:
\begin{equation}\label{Eq:shared_ss}
    \bigl(\mathcal{D}_R(\varepsilon) -
    \mathcal{D}_L(\varepsilon)\bigr)\rho_{ss} = 0.
\end{equation}
From Eq. \ref{Eq:shared_ss}, we trivially conclude that the superoperators $\mathcal{D}_R(\varepsilon)$ and $\mathcal{D}_L(\varepsilon)$ act in the same way on $\rho_{ss}$: $\mathcal{D}_R(\varepsilon)\rho_{ss} = \mathcal{D}_L(\varepsilon)\rho_{ss}$. It directly implies that the $\mathcal{L}(s_k, \varepsilon)$ action on $\rho_{ss}$ is the same regardless of any input values $s_k$. Consequently, we have that
\begin{equation*}
   \mathcal{L}(s_1, \varepsilon)\rho_{ss} =
   \mathcal{L}(s_2, \varepsilon)\rho_{ss} =
   \mathcal{L}(1/2, \varepsilon)\rho_{ss} = 0.
\end{equation*}
We can readily check that the unique stationary stater of $\mathcal{L}(1/2, \varepsilon)$ is the fully mixed state  $\mathbb{I}_{N_{b}}/\text{dim}(\mathbb{H}_{N_{b}})$ and, consequently, the same must hold for the stationary states corresponding to the generic input values $s_1$ and $s_2$. However, under the condition $0 \leq \varepsilon < 1$, we can verify, from a direct calculation, that $\mathbb{I}_{N_{b}}/\text{dim}(\mathbb{H}_{N_{b}})$ is the stationary state that only corresponds to the input value $s_k = 1/2$. 

Consequently, we can conclude that two different input values cannot share the same stationary state. $\blacksquare$

\subsection{Input separability and universality}

Theorem 4 proves that when $\varepsilon \neq 1$, for two different input values, the corresponding stationary states of $\mathcal{L}(s_k, \varepsilon)$ will be distinct. This result, as explained in \cite{Sannia2024}, directly implies that each of the models described by Eq. 4 of the main text, when the boundary conditions are not periodic, can separate different input sequences. Going more into detail, for two distinct series $\bm{s}_1$ and $\bm{s}_2$ belonging to $K^{-}([0,1])$, there exists a specific value of $\Delta t$ such that the corresponding reservoir states, after the injection of these two series, will differ, thereby allowing for their discrimination. This property is known as input separability. 

To conclude, the validity of this last feature implies that any model described by Eq. 4 of the main text, when $0 \leq \varepsilon < 1$, forms a universal class of reservoir computers \cite{Sannia2024}. Universality, in this case, means that any fading memory map, i.e. a functional acting on $K^{-}([0,1])$ which is continuous under the norm defined above, can be arbitrarily well approximated combining outputs of the reservoir. 

\section{Performance evaluation methodology}\label{Sec:perfmet}

In this section, we will show the details of the strategy used for evaluating the performance of all the considered reservoir models. We first recall that, to give representative results that do not depend on statistical fluctuations, we always computed the capacity over different realizations. In particular, for each considered task, the performance values have been obtained by averaging over 100 different random combinations of initial conditions, input sequences, and on-site disorder realizations. Moreover, for each of these realizations, we first let the system evolve for $1000$ time steps, according to the particular series of interest, so that its state will be independent of its initial condition (washout phase), as a consequence of the echo state property \cite{mujal2022time}. After that, $1000$ data points were utilized for training the free weights of the output layer (training phase) \cite{BookRC}, and, finally, other $1000$ points were dedicated to evaluating the performance (test phase). Through numerical verification, we ensured that the datasets used in all the tasks considered were sufficiently large to prevent over-fitting, allow the collection of meaningful statistics for accurate performance evaluations, and let the performance be initial state independent. At each time step, to fully exploit the reservoir capabilities,  we have always taken a linear combination of all the density matrix elements as the output layer. In addition, to model the effect of shot noise in the performance, for a number of available measurements equal to $N_s$, we followed the results of \cite{mujal2022time}. In particular, assuming a large ensemble with $N_s \gg 1$,  the statistical error can be modeled by adding a zero mean Gaussian noise term to each output layer element, where we considered the standard deviation to be the maximum possible: $\sigma_{max}(N_s) = 1/\sqrt{N_s}$. Following the established notation, the reservoir outputs $x_k$ are given by:
\begin{align*}
    x_k = \sum_{i,j}\alpha_{i,j}^R \mathcal{N}\big(\Re(\rho_{k})_{i,j} , \sigma_{max}(N_s)\big) + \sum_{i,j}\alpha_{i,j}^I \mathcal{N}\big(\Im(\rho_{k})_{i,j}, \sigma_{max}(N_s)\big) + \alpha_0,
\end{align*}
where the summations iterate over all the density matrix elements. Here, $\Re(\rho_{k})$ and $\Im(\rho_{k})$ are the real and imaginary parts of $\rho_k$, respectively; $\mathcal{N}(\mu, \sigma)$ denotes a sample from a Gaussian distribution with mean $\mu$ and standard deviation $\sigma$; $\alpha_{i,j}^R$, $\alpha_{i,j}^I$ and $\alpha_0$ are the free weights to be optimized for any given QRC task.

Using the $scikit$-$learn$ library \cite{scikit}, for each task considered, we determined the optimal weights by minimizing the following least squares function during the training phase:
\begin{align*}
    \sum_{k}(x_k-y_k)^2,
\end{align*}
where $k$ iterates over the training steps and the $y_k$ values refer to the target series of interest.

\end{document}